\documentclass[aps,showpacs,twocolumn,prd]{revtex4}
\usepackage{graphicx}
\usepackage{amssymb}
\usepackage{amsmath}
\newcommand{\be}{\begin{equation}}
\newcommand{\ee}{\end{equation}}
\newcommand{\ben}{\begin{eqnarray}}
\newcommand{\een}{\end{eqnarray}}
\newcommand{\bes}{\begin{subequations}}
\newcommand{\ees}{\end{subequations}}
\newcommand{\bb}{\bibitem}

\newcommand{\sech}{{\rm sech}}
\begin{document}
\title{New family of sine-Gordon models}
\author{D. Bazeia, L. Losano, R. Menezes and M.A.M. Souza}
\affiliation{Departamento de F\'\i sica, Universidade Federal da Para\'\i ba, 58051-900 Jo\~ao Pessoa, Para\'\i ba, Brazil}

\begin{abstract}
This work deals with a new family of models, which includes the sine-Gordon model and the double-sine-Gordon, triple-sine-Gordon and so on.
The investigation is based on a deformation procedure, which is used to deform a well-known model, to get to the family of sine-Gordon models. Due to properties of the procedure, we get to the models and find the corresponding solutions explicitly, together with all the important features they engender.
\end{abstract}
\maketitle

The sine-Gordon model is described by a relativistic real scalar field in $(1,1)$ spacetime dimensions, and is of great interest in a diversity of contexts. It is a paradigm for integrable models, and it has been studied in a diversity of scenarios, as one can see in  \cite{Rub} and in more recent years in the works \cite{sgO,sgA}. The double and triple sine-Gordon model are also of interest, but they are harder to solve \cite{dsg,ref9}.

The solutions that we search for are static classical configurations which solve equations of motion of relativistic models described by a single real scalar field $\phi$ in $(1,1)$ spacetime dimensions. Here we deal mainly with the presence of solutions of the topological type, which solve first-order differential equations. Such solutions are Bogomol'nyi-Prasad-Sommerfield (BPS) states \cite{BPS}, and in general they are stable under small fluctuations \cite{BS}. They are minima energy configurations, and may play important role in applications in several distinct areas such as high energy physics \cite{MS,V} and condensed matter, where they can be used in several different contexts, in particular to describe pattern formation in nature \cite{W}. 

In this work we study a family of models which includes the sine-Gordon model as the first in a sequence of sine-Gordon-like models. The set of models represents a new family, which appears very naturally within the context of the deformation procedure set forward in \cite{BLM}. Applications of current interest
in high energy physics concerns the use of such models within the Randall-Sundrum braneworld context \cite{RS}; see also \cite{F,brane} and references therein, which explore how the scalar field may contribute to smooth the brane profile in the five-dimensional scenario with a single extra dimension of infinite extent. 

The investigation starts with models described by the Lagrange density \cite{BLM}
\be\label{equ1}
{\cal L}=\frac12\partial_\mu\chi\partial^\mu\chi-V(\chi)
\ee
where $V(\chi)$ is the potential which specifies the model under consideration, and $\chi$ is a real scalar field . We focus mainly on the general properties of the models and their corresponding solutions; a more detailed work including applications is under preparation. For the model \eqref{equ1}, the equation of motion for a generic static field $\chi=\chi(x)$ is given by ${d^{2}\chi}/{dx^2}={dV}/{d\chi}$. To search for topological solutions, we write the potential in the form 
\be\label{equ3}
V(\chi)=\frac{1}{2}W_{\chi}^2
\ee
where $W=W(\chi)$ is a smooth function of the field, with $W_{\chi}=dW/d\chi$. In this case the equation of motion becomes
\be\label{equ4}
\frac{d^2\chi}{dx^2}=W_{\chi}W_{\chi\chi}
\ee
This equation is solved with solutions of the first-order equation
\be\label{equ5}
\frac{d\chi}{dx}=W_{\chi}
\ee
Since the potential does not see the sign of $W$, there is another equation associated to the above equation \eqref{equ5}, obtained by changing $W$ to $-W$. 

A topological solution which solves the first-order equation has energy minimized to the value $E=|\Delta W|=|W(\chi(\infty))-W(\chi(-\infty))|$.
This is known as the Bogomol'nyi bound \cite{BPS}, and it shows that we can get the energy of the solutions without knowing the solutions explicitly.

Before introducing the family of models, let us focus attention on the deformation procedure set forward in \cite{BLM}, and further used in \cite{other} in different contexts. For instance, in the first work in \cite{other} one used the deformation procedure to get to the double sine-Gordon model, and there we highlighted an important advantage of the deformation procedure, which relies on finding the topological solutions analytically.
To use the deformation procedure, we firstly introduce the standard model 
\be\label{phi4}
V(\chi)=\frac12{(1-\chi^2)^2}
\ee
which is the $\chi^{4}$ model with spontaneous symmetry breaking. $V(\chi)$ is the potential of the starting model. Here we are using natural units, and we have rescaled the field, and the space and time coordinates to make them dimensionless. This model has as topological defects the BPS states $\chi_{\pm}(x)=\pm\tanh(x)$, where we are taking the center of the solutions at the origin, for simplicity. The energy density is given by
$\varepsilon(x)={\rm sech}^4(x)$, which gives the energy $E=4/3$. We also note that the potential has minima at ${\bar\chi}_\pm=\pm1$ which obeys $V({\bar\chi}_\pm)=0,$ and $V^\prime({\bar\chi}_\pm)=0$, where $V^\prime(\chi)=dV/d\chi$. 

According to the deformation procedure, we can consider another model described by $U(\phi)$, which is written in terms of the starting model $V(\chi)$ as \cite{BLM} 
\be\label{equ7}
U(\phi)=\frac{V(\chi\to f(\phi))}{f^{\prime 2}(\phi)}
\ee
where $f(\phi)$ is the deformation function, and $\phi$ is the new field. $U(\phi)$ is the potential of the deformed model. In this case, if $\chi(x)$ is a static solution of the starting model, then we get that $\phi(x)=f^{-1}(\chi(x))$ is a solution of the new, deformed model. If we use this procedure to the model \eqref{phi4} we have
\be\label{equ8}
U(\phi)=\frac12\frac{(1-f^2(\phi))^2}{f^{\prime 2}(\phi)}
\ee
We note that the model \eqref{phi4} has an interesting property, which can be proven easily: if $f(\phi)$ is the function to be used to deform the model,
then $1/f(\phi)$ is another deformation function, which gives the very same deformed model. We can then say that for the model \eqref{phi4}, both
\be\label{P}
f(\phi)\;\; {\rm and}\;\;\frac1{f(\phi)}
\ee
form a pair of deformations which lead to the very same model. This property will be used below to introduce the new family of sine-Gordon models.

Let us now choose the model \eqref{phi4} and the deformation function in the form $f(\phi)=\tan(\phi)$. This leads to the model defined by
\be\label{equ14}
W(\phi)=\frac12\sin(2\phi),\;\;\;\;\; V(\phi)=\frac12{\cos^2(2\phi)}
\ee
In this case the deformation function depends only on $\phi$, and the corresponding deformed model describes the sine-Gordon model, with no extra parameter involved in the procedure. Here we can label the minima and maxima of the potential according to
\bes\ben
{\phi}_{\rm min}^n=\pm\frac{n}4\pi,\;\;\; {\rm for}\;\;\;n=1,3,5,...
\\
{\phi}_{\rm max}^n=\pm\frac{n}4\pi,\;\;\; {\rm for}\;\;\;n=0,2,4,...
\een\ees
This model has solutions which can be obtained from the inverse of $f(\phi)=\tan(\phi)$ and $1/f(\phi)=\cot(\phi)$; they are described by, respectively 
\bes\label{ssg}\ben
\phi^1_m(x)&=&\pm\arctan(\tanh(x))\pm m\pi
\\
\phi^2_m(x)&=&\pm{\rm arccot}(\tanh(x))\pm m\pi
\een\ees
Here $m=0, 1, 2,...$ identifies the particular sector of the sine-Gordon model, which has an infinity of topological sectors. We then see that we go from the $\chi^4$ model, which contains a single sector, to the sine-Gordon model, which contains an infinity of sectors, with the use of the deformations $f(\phi)=\tan(\phi)$ and $1/f(\phi)=\cot(\phi)$, which are periodic functions.

In the sine-Gordon model, the energy density of the BPS states can be written in the form
\be\label{edsg}
\epsilon(x)={\rm sech}^2(2x)
\ee
and so all the topological sectors have the same energy, $E=1$. It is interesting to note that although the starting model has a single topological sector, with kink and antikink solutions $\pm\tanh(x)$, the deformed model has an infinity of topological sectors, and we get to the corresponding solutions with $\tan(\phi)$ and $\cot(\phi)$, with the integer $m$ (which naturally appears from the deformation function) mapping each one of the many individual sectors of the model.

To show how to generate the new family of sine-Gordon models, let us consider another deformation function, $f_r(\phi)=r\tan(\phi)$, with $r$ real and positive, $r\in(0,\infty)$. Together we also have $1/f_r(\phi)=(1/r)\cot(\phi)$. Here we get to the double sine-Gordon model, with
\be
W(\phi)=\frac12\left(\frac1r-r\right)\phi+\frac14\left(\frac1r+r\right)\sin(2\phi)
\ee
and 
\be\label{dsg}
V(\phi)=\frac{1}{2r^2}\left((1+r^2)\cos^2(\phi)-r^2\right)^2
\ee 
This model depends on the parameter $r$, which engenders the interesting feature of controlling the two distinct sectors of the model. We note that the limit $r\to1$ leads us back to the former sine-Gordon model, and $r$ and $1/r$ exchange the two distinct sectors of the model.

In the double sine-Gordon model, the minima of the potential are given by
\be
\phi^m_{\rm min}=\pm\arctan(1/r)\pm m\pi,\;\;m=0, 1, 2,...
\ee
and the maxima are 
\be
\phi^m_{\rm max}=\pm\frac{m}{2}\pi,\;\;m=0, 1, 2,....
\ee
We note that the minima depends on $r$, but this is not the case for the maxima, which have fixed positions. By the way, we see that the position of the higher and lower maxima exchange place when $r$ changes to $1/r$. This can be seem from the height of the maxima, which are given by
\be
h_1(r)=\frac1{2r^2},\;\;\;\;\;h_2(r)=\frac{r^2}{2}
\ee
They exchange place when one changes $r\to1/r$. The double sine-Gordon model has two families of distinct topological sectors, which we label with $a$ and $b$, so the topological solutions of the potential described by \eqref{dsg} has to be characterized by two distinct solutions, $\phi_a(x)$ and $\phi_b(x)$. They are obtained by the two deformation functions $f(\phi)$ and $1/f(\phi)$. For instance, with $f(\phi)=r\tan(\phi)$ we obtain 
\be\label{bsgl}
\phi_{1}(x)=\pm\arctan\left(\frac{1}{r}\tanh(x)\right)\pm m\pi
\ee
with $m=0, 1, 2,...$ identifying one of the two distinct sectors of the model. To obtain the other solutions, describing the other family of sectors of the model, we use the function $1/f_r(\phi)=(1/r)\cot(\phi)$. It leads to the same potential, but it maps the solutions of the starting $\chi^4$ model into other defects, which are given explicitly by
\be\label{dsgs}
\phi_{2}(x)=\pm{\rm arccot}\left(r \tanh(x)\right)\pm m\pi
\ee
In Fig.~1 we plot the potentials of the $\chi^4$ and the double sine-Gordon models. We use $r=2$, and take $\phi$ in the interval $[-\arctan(1/r),\pi-\arctan(1/r)]$, which shows the sectors corresponding to the two solutions given by Eqs.~\eqref{bsgl} and \eqref{dsgs}, in the case $m=0$. In this Fig.~1 we also illustrate how the deformation functions $f_r(\phi)$ and $1/f_r(\phi)$ map the two family of topological sectors of the model.

\begin{figure}[t]
\includegraphics[scale=0.5,width=5cm]{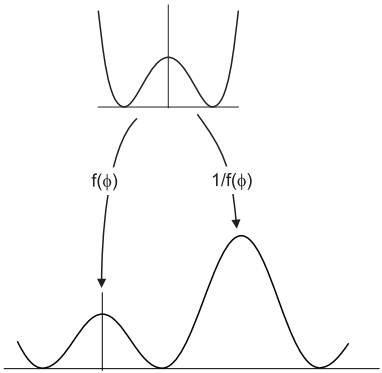}
\caption{Plot of the potential of the double sine-Gordon model for $r=2$, and for $\phi$ in the interval $[-\arctan(1/r),\pi-\arctan(1/r)]$. The illustration shows how the functions act to map the topological sector of the starting model into the two distinct family of topological sectors of the deformed model.} 
\end{figure}

The energy densities of the deformed solutions are given by
\be\label{equ33}
\varepsilon^1(x)=\frac{r^2\sech^4(x)}{(r^2+\tanh^2(x))^2}
\ee
\be\label{equ34}
\varepsilon^2(x)=\frac{r^2\sech^4(x)}{\left(1+r^2\tanh^2(x)\right)^2}
\ee
and the total energy of the BPS states can be calculated straightforwardly. We note that in the limit $r\to1$ one gets $\epsilon^1(x)=\epsilon^2(x)$, which then exactly reproduces the result \eqref{edsg} of the sine-Gordon model.

We go on introducing another deformation function, given by
\be\label{def1}
f_{rs}=\tan\{s\;\arctan[r\tan(\phi)]\}
\ee
where $r$ is real and positive, $r\in(0,1)$ or $r\in(1,\infty)$, with $r\neq1$, and $s$ is positive integer, $s=1,2,3,...$. With the property \eqref{P} we can introduce three other deformations
\be\label{def2}
1/f_{rs}=\cot\{s\;\arctan[r\tan(\phi)]\}
\ee
and 
\be\label{def3}
g_{rs}=\tan\{s\;\arctan[(1/r)\cot(\phi)]\}
\ee
and
\be\label{def4}
1/g_{rs}=\cot\{s\;\arctan[(1/r)\cot(\phi)]\}
\ee
These deformations lead us to the same potential
\ben\label{v12}
V(\phi)&=&\frac{1}{2r^2s^2}\{2\cos^2[s\;\arctan(r\tan(\phi))]-1\}^2\nonumber
\\
&&[(1-r^2)\cos^2(\phi)+r^2]^2
\een
Now we have four functions, but the parameter $s$ is very interesting and can be used to lead to new family of models. We see that for $s=1$ we get to the double sine-Gordon model, which contains two ($s+1$, with $s=1$) distinct topological sectors, labeled by $1$ and $2$. As we are going to show, for $s=2$ we get to the triple sine-Gordon model, which contains three ($s+1$, with $s=2$) distinct topological sectors, labeled by $1$, $2$, and $3$. For $s=3$ we get to the quadruple sine-Gordon model and so on. The other parameter, $r$, plays the same role as before, and it controls the position of the minima and the height of the maxima.

The minima of the general model are given by
\be
\phi^{n,m}_{\rm min}\!\!\!=\!\!\pm\arctan\left(\frac1r\tan\left((2n\!-\!1)\frac{\pi}{4s}\right)\right)\!\pm\!m\pi\;\;\;
\ee
with $n$ integer, $1\leq n\leq s$, and $m=0, 1, 2,...\,$.

To classify the height of the maxima we proceed as follows: for $0\leq\phi\leq\pi/2$ we have $l=s+1$ maxima, with the corresponding height increasing for $r\in(1,\infty)$, and decreasing for $r\in(0,1)$.  The first is at $\phi^1_{max}=0$ and the last one at $\phi^{s+1}_{max}=\pi/2$, with the heights
\ben\label{hgen1}
h_1=\frac{1}{2r^2s^2},\;\;\;h_{s+1}=\frac{r^2}{2s^2},
\een
respectively. The other $s-1$ maxima are all in between the above two maxima; they are located at
\be\label{hs}
\phi^l_{max}=\frac12\left(\phi^{l,0}_{min}-\phi^{l-1,0}_{min}\right)
\ee
for $l=2,3,...,s$. The height of these maxima can be calculated for each specific value of $s$, with the use of \eqref{hs} and the potential \eqref{v12}.

\begin{figure}[t]
\includegraphics[scale=0.5,width=5cm]{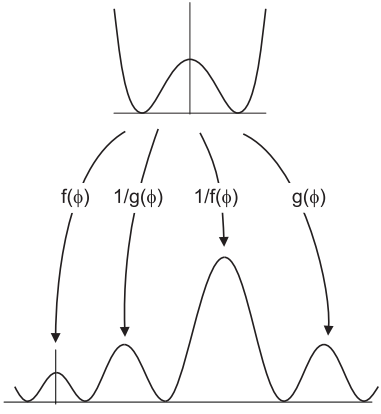}
\caption{Plot of the potential of the triple sine-Gordon model for $r=2$ and for $\phi$ in the interval $[-\arctan((1/r)\tan(\pi/8)),\pi-\arctan((1/r)\tan(\pi/8))]$. The illustration also shows how the functions act to map the topological sector of the starting model into the three distinct family of topological sectors of the deformed model.} 
\end{figure}

The topological solutions are given by, for a given integer $s$
\bes\label{alls}\ben
\phi^{1,m}_{k,s}&=&\arctan\left(\frac1r\tan\left(F_{k,s}\right)\right)\pm m\pi
\\
\phi^{2,m}_{k,s}&=&\arctan\left(r\tan\left(F_{k,s}\right)\right)\pm\left(m+\frac12\right)\pi
\\
\phi^{3,m}_{k,s}&=&\arctan\left(\frac1r\tan\left(G_{k,s}\right)\right)\pm m\pi
\\
\phi^{4,m}_{k,s}&=&\arctan\left(r\tan\left(G_{k,s}\right)\right)\pm\left(m+\frac12\right)\pi
\een\ees
where $m=0, 1, 2,...,$ $k=0,1,....,$ ($k$ is restricted by $s$; see below) and 
\bes\label{with}\ben
F_{k,s}&=&\frac1s\arctan\chi(x)+k\frac{\pi}{s}
\\
G_{k,s}&=&\frac1s\arctan\chi(x)+\left(k+\frac12\right)\frac{\pi}{s}
\een\ees
with $\chi(x)=\pm\tanh(x)$ giving rise to the defect and anti-defect solutions. In this case, we cannot express the energy density analytically for $s$ arbitrary, but this can be done for every integer $s$. 

To see how the family of models behaves, let us note that the periodicity of the potential is $\pi$, and for $m=0$ all the above solutions
are inside the interval
\[
-\arctan\left(\frac1r\tan\left(\frac{\pi}{4s}\right)\right)\leq\phi\leq\pi-\arctan\left(\frac1r\tan\left(\frac{\pi}{4s}\right)\right)
\]
In the general case, for $s=1,2,...,$ we have that the above interval of periodicity of the model is divided in $2s$ parts. For $s=1$ and $r=1$ we get to the sine-Gordon model, and for $s=1$ and $r$ arbitrary, we get to the double sine-Gordon model, as we have already seen explicitly. For $s>1$ the solutions are given by: for $s=2,4,...,$ even, we have
\be\label{even}
\phi^{1,0}_{k,s},\;\phi^{2,0}_{k,s},\;\phi^{3,0}_{k,s},\;\phi^{4,0}_{k,s},\;\; {\rm for}\;k=0,1,...,\frac{s-2}{2}
\ee 
and for $s=3,5,...,$ odd, we get
\bes\label{odd}\ben
&&\phi^{1,0}_{k,s},\;\phi^{2,0}_{k,s},\;\phi^{3,0}_{k,s},\;\phi^{4,0}_{k,s},\;\; {\rm for}\;k=0,1,...,\frac{s-3}{2}\;\;\;
\\
&&\phi^{1,0}_{k,s},\;\phi^{4,0}_{k,s},\;\; {\rm for}\;k=\frac{s-1}{2}
\een\ees
The above equations \eqref{alls}, \eqref{with}, \eqref{even}, and \eqref{odd} give all the solutions of each member of the family of sine-Gordon models.

We illustrate the case $s=2$ in Fig.~2, where we show how the deformation functions act to map the three families of solutions of the triple sine-Gordon model.
Here we have 
\be
W(\phi)=\alpha\phi+\beta \sin(2\phi)+\gamma \arctan[r\tan(\phi)]
\ee
where
\bes\ben
\alpha&=&\frac{(1+r^2)(1-10r^2+r^4)}{2(1-r^2)^2}
\\
\beta&=&\frac{1+6r^2+r^4}{4(1-r^2)};\;\;\;\gamma=\frac{8r^3}{(1-r^2)^2}
\een\ees
The height of the first and third maxima are given according to \eqref{hgen1}, and the height of the second, intermediate maximum has the form 
\be
h_2=\frac{4r^2[\sqrt{1+6r^2+r^4}-\sqrt{2}(1+r^2)]^2}{(1-r^2)^4}
\ee
\begin{figure}[t]
\includegraphics[scale=0.5,width=6cm]{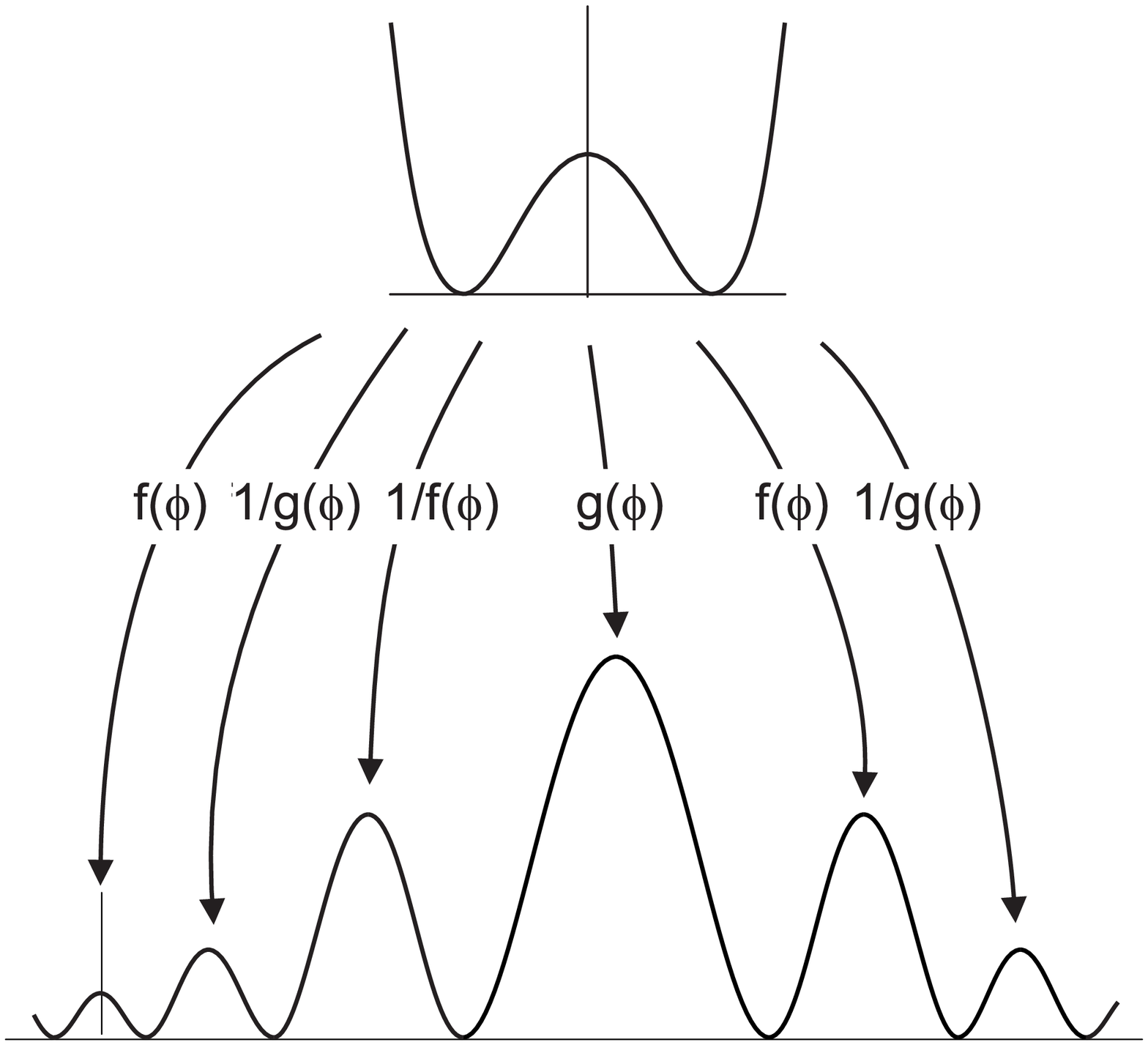}
\caption{Plot of the potential of the quadruple sine-Gordon model for $r=2$ and for $\phi$ in the interval $[-\arctan((1/r)\tan(\pi/12)),\pi\!-\!\arctan((1/r)\tan(\pi/12))]$. The illustration shows how the functions act to map the topological sector of the starting model into the four distinct family of topological sectors of the deformed model.} 
\end{figure}
The energy densities of the three solutions corresponding to the three distinct topological sectors are given by, respectively:
\bes\ben
\epsilon^1(x)&=&\frac{r^2{\rm sech}^2(2x)}{4[(1-r^2)\cos^2\theta(x)-1]^2}
\\
\epsilon^2(x)&=&\frac{r^2{\rm sech}^2(2x)}{4[(1-r^2)\cos^2(\theta(x)+\pi/4)-1]^2}
\\
\epsilon^3(x)&=&\frac{r^2{\rm sech}^2(2x)}{4[(1-r^2)\cos^2\theta(x)+r^2]^2}
\een\ees 
where $\theta(x)={\rm arctan}[\tanh(x)]$. Here we are using $1,2$, and $3$ to label each one of the three distinct topological sectors of the triple sine-Gordon model. The total energy corresponding to each one of the topological sectors can be calculated straightforwardly.

We further illustrate the procedure with the quadruple sine-Gordon model. In this case, we use  $s=3$ and in Fig.~3 we show the potential and illustrate how the solutions are mapped as they appear from \eqref{odd} after taking $s=3$, with the four distinct topological sectors. The procedure works nicely, and all the other cases (with $s=4,5,...$) follow naturally. In particular, if we introduce another parameter [multiplying the functions \eqref{def1} and \eqref{def3}] we can then exchange the position and height of the maxima, making the models more general. This and other related issues will be explored elsewhere.

In summary, we have used the deformation procedure introduced in \cite{BLM} to build a new family of models of the sine-Gordon type. The models start with the sine-Gordon model itself, and include the double sine-Gordon model, and the triple, quadruple and so on. They are controlled by the two real parameters, $r$ and $s$: $r$ adjusts the position and height of the maxima, and $s$ identifies each one of the members of the proposed family, controlling the number of distinct topological sectors of the model. The procedure allows to build the models and their explicit solutions, leading to new models and exposing all the important features of the topological solutions.

The authors would like to thank CAPES, CNPq and PRONEX-CNPq-FAPESQ for partial support.


\end{document}